\documentclass[12pt]{article}
\usepackage{epsfig,psfrag}
\catcode`@=11

\def\@citex[#1]#2{\if@filesw\immediate\write\@auxout{\string\citation{#2}}\fi
  \def\@citea{}\@cite{\@for\@citeb:=#2\do
    {\@citea\def\@citea{,\penalty\@m}\@ifundefined
      {b@\@citeb}{{\bf ?}\@warning
       {Citation `\@citeb' on page \thepage \space undefined}}%
\hbox{\csname b@\@citeb\endcsname}}}{#1}}

\def\citer{\@ifnextchar [{\@tempswatrue\@citexr}{\@tempswafalse\@citexr[]}}

\def\@citexr[#1]#2{\if@filesw\immediate\write\@auxout{\string\citation{#2}}\fi
  \def\@citea{}\@cite{\@for\@citeb:=#2\do
    {\@citea\def\@citea{--\penalty\@m}\@ifundefined
       {b@\@citeb}{{\bf ?}\@warning
       {Citation `\@citeb' on page \thepage \space undefined}}%
\hbox{\csname b@\@citeb\endcsname}}}{#1}}

\begin{document}
 
\thispagestyle{empty}
\begin{flushright}
LMU 24/03\\
May 2004
\end{flushright}

\vspace*{1.5cm}
\centerline{\Large\bf CP Violation in $B \to\pi^+\pi^-$}
\vspace*{0.3cm}
\centerline{\Large\bf and the Unitarity Triangle}
\vspace*{2cm}
\centerline{{\sc Gerhard Buchalla\footnote{
              buchalla@theorie.physik.uni-muenchen.de}} and 
{\sc A. Salim Safir\footnote{safir@theorie.physik.uni-muenchen.de}}}
\bigskip
\bigskip
\centerline{\sl Ludwig-Maximilians-Universit\"at M\"unchen, 
Sektion Physik,} 
\smallskip
\centerline{\sl Theresienstra\ss e 37, D-80333 Munich, Germany}

\vspace*{0.5cm}
\begin{abstract}
We analyze the extraction of weak phases from CP violation in 
$B\to\pi^+\pi^-$ decays. We propose to determine the unitarity triangle 
$(\bar\rho,\bar\eta)$ by combining the information on mixing induced CP 
violation in $B\to\pi^+\pi^-$, $S$, with the precision observable $\sin 2\beta$
obtained from the CP asymmetry in $B\to\psi K_S$. It is then possible to write 
down exact analytical expressions for $\bar\rho$ and $\bar\eta$ as simple
functions of
the observables $S$ and $\sin 2\beta$, and of the penguin parameters $r$ and 
$\phi$. As an application clean lower bounds on $\bar\eta$ and
$1-\bar\rho$ can be derived as functions of $S$ and $\sin 2\beta$,
essentially without hadronic uncertainty. Computing $r$ and $\phi$
within QCD factorization yields precise determinations of $\bar\rho$ and 
$\bar\eta$ since the dependence on $r$ and $\phi$ is rather weak.
It is emphasized that the sensitivity to the phase $\phi$ enters only 
at second order and is extremely small for moderate values of this 
phase, predicted in the heavy-quark limit. 
Transparent analytical formulas are further given and discussed for the 
parameter $C$ of direct CP violation in $B\to\pi^+\pi^-$. We also discuss 
alternative ways to analyze $S$ and $C$ that can be useful if new physics 
affects $B_d$--$\bar B_d$ mixing. Predictions and uncertainties for $r$ and 
$\phi$ in QCD factorization are examined in detail. It is pointed out that a 
simultaneous expansion in $1/m_b$ and $1/N$ leads to interesting 
simplifications. At first order infrared divergences are absent, while the 
most important effects are retained. Independent experimental tests of the 
factorization framework are briefly discussed. 
\vspace*{0.3cm}


\end{abstract}


\newpage
\pagenumbering{arabic}

\section{Introduction}\label{sec:intro}

The main goal of the current experimental program at the SLAC
and KEK $B$-meson factories is a stringent test of the standard model 
description of CP violation. In the future this aim will be pursued
with measurements of still higher precision from hadron machines at
Fermilab and CERN. A crucial benchmark is the time-dependent CP violation
in $B\to\psi K_S$ decays, which allows us to infer the CKM phase $\beta$
with negligible hadronic uncertainties.
Likewise of central importance for obtaining additional information on
CKM parameters is the time-dependent CP violation,
both mixing-induced ($S$) and direct ($C$), in $B\to\pi^+\pi^-$.
However, in this case the extraction of weak phases is complicated
by a penguin component in the decay amplitude, which carries a weak phase
different from the leading tree-level contribution. This leads to a 
dependence of the CP asymmetries in $B\to\pi^+\pi^-$ on hadronic physics
and to corresponding theoretical uncertainties.
A possible strategy to circumvent this problem is the isospin analysis
by Gronau and London \cite{GL}, where also the branching ratios of
$B^+\to\pi^+\pi^0$, $B\to\pi^0\pi^0$ and their charge conjugates have
to be measured. This method is theoretically very clean, but
the difficulty to measure $B\to\pi^0\pi^0$ decays
with sufficient accuracy and the existence of discrete ambiguities
are likely to prevent a successful realization.

It is the purpose of this paper to demonstrate how the information
on weak phases contained in the CP asymmetries of $B\to\pi^+\pi^-$
itself can be extracted in an optimal way.
To some extent theoretical input on the penguin-to-tree ratio will
be needed and can be provided by the QCD factorization approach
\cite{BBNS1,BBNS2,BBNS3}.
However, we will show that the impact of uncertainties in the
calculation is in fact very mild. Moreover, even if the detailed
predictions of QCD factorization are ignored, it is still possible to 
derive rigorous bounds on the CKM unitarity triangle, using only
very conservative assumptions.

In order to derive these results we propose the following strategy.
First, the time-dependent CP asymmetries in $B\to\pi^+\pi^-$ are
expressed in terms of Wolfenstein parameters $\bar\rho$ and $\bar\eta$.
At the same time purely hadronic quantities are systematically isolated
from CKM parameters, introducing magnitude $r$ and phase $\phi$ of
a suitably normalized penguin-to-tree-ratio \cite{BBNS3}. We then combine the
observable $S(\bar\rho,\bar\eta,r,\phi)$ with the accurately known
value of $\sin 2\beta(\bar\rho,\bar\eta)$ from $B\to\psi K_S$.
This allows us to obtain the exact unitarity triangle, $\bar\rho$ and
$\bar\eta$, in a simple analytical form, depending only on
$\sin 2\beta$, $S$ and the hadronic quantities $r$ and $\phi$.
The dependence on the latter turns out to be particularly transparent,
which greatly facilitates any further analysis.
We are then able to derive bounds on the unitarity triangle practically
free of hadronic uncertainties, or to fix $\bar\rho$ and $\bar\eta$
with theoretical input for $r$ and $\phi$.

There is already an extensive literature on the subject of
extracting information on weak mixing angles from CP violation
in $B\to\pi^+\pi^-$ \citer{SW,ALP}. In these papers
important aspects of the problem have been discussed and suggestion
were made to constrain theoretical uncertainties.
Here we present a new way of exploiting the information
contained in the CP violation observables $S$ and $\sin 2\beta$.
The crucial elements are a 
definition of hadronic quantities $r$ and $\phi$ independent of
CKM parameters, the direct formulation of weak phases in terms
of the basic Wolfenstein parameters $\bar\rho$, $\bar\eta$, 
the resulting analytical determination of the unitarity triangle
and the exact, explicit and very simple dependence on $r$ and $\phi$.
This in turn greatly facilitates the analysis of theoretical uncertainties
and gives, in combination with results based on the heavy-quark limit,
robust determinations of the unitarity triangle, or rigorous CKM bounds 
with minimal hadronic input.
These ideas were first presented in \cite{BS}. 
Subsequently, this analysis has been further discussed
by Botella and Silva \cite{BSI} and Lavoura \cite{LAV}.

The paper is organized as follows.
In sec. 2 we collect important basic formulas describing CP violation
in $B\to\pi^+\pi^-$.
In sec. 3 we discuss the theory of the penguin parameters $r$ and $\phi$
in the framework of QCD factorization. Based on previous work we
address in particular the issue of theoretical uncertainties.
In addition we investigate the analysis of $B\to\pi^+\pi^-$ decay amplitudes
in a simultaneous expansion in both $1/m_b$ and $1/N$, where $N$ is the
number of colours. An interesting pattern of systematic simplifications
resulting from the double expansion is pointed out.
After this discussion of the hadronic input, we turn to our
phenomenological analysis. Sec. 4 explores the determination of the
unitarity triangle from $S$ and $\sin 2\beta$ within the standard model.
Simple analytical expressions are presented and theoretically clean
bounds are derived. We also evaluate the standard model expectation
for $S$ using results from QCD factorization. Sect. 5 considers
the determination of CKM angle $\gamma$ from $S$, $\sin 2\beta$ and
$|V_{ub}/V_{cb}|$. This alternative possibility is useful if
new physics affects the phase of $B_d$--$\bar B_d$ mixing.
Sec. 6 examines what can be learned from $C$, the parameter of
direct CP violation in $B\to\pi^+\pi^-$. Methods to validate the predictions
of QCD factorization for $B\to\pi^+\pi^-$ using additional observables
are described in sec. 7. We summarize our main results in sec. 8.

\section{Basic Formulas}

The time-dependent CP asymmetry in $B\to\pi^+\pi^-$ decays
is defined by
\begin{eqnarray}\label{acppipi}
A^{\pi\pi}_{CP}(t) &=& 
\frac{B(B(t)\to\pi^+\pi^-)-B(\bar B(t)\to\pi^+\pi^-)}{
  B(B(t)\to\pi^+\pi^-)+B(\bar B(t)\to\pi^+\pi^-)} \nonumber \\
&=& - S\, \sin(\Delta m_B t) + C\, \cos(\Delta m_B t)
\end{eqnarray}
where
\begin{equation}\label{scxi}
S=\frac{2\, {\rm Im}\xi}{1+|\xi|^2}\qquad
C=\frac{1-|\xi|^2}{1+|\xi|^2}\qquad
\xi=e^{-2 i\beta}\,\frac{e^{-i\gamma}+P/T}{e^{+i\gamma}+P/T}
\end{equation}
In terms of the Wolfenstein parameters $\bar\rho$ and $\bar\eta$
\cite{WO,BLO} the CKM phase factors read
\begin{equation}\label{gambetre}
e^{\pm i\gamma}=\frac{\bar\rho\pm i \bar\eta}{\sqrt{\bar\rho^2+\bar\eta^2}}
\qquad
e^{-2 i\beta}=\frac{(1-\bar\rho)^2 -\bar\eta^2 - 2 i\bar\eta(1-\bar\rho)}{
                     (1-\bar\rho)^2 + \bar\eta^2}
\end{equation}
The penguin-to-tree ratio $P/T$ can be written as
\begin{equation}\label{ptrphi}
\frac{P}{T}=\frac{r e^{i\phi}}{\sqrt{\bar\rho^2+\bar\eta^2}}
\end{equation}
The real parameters $r$ and $\phi$ defined in this way are
pure strong interaction quantities without further dependence
on CKM variables.

For any given values of $r$ and $\phi$ a measurement of $S$ 
defines a curve in the ($\bar\rho$, $\bar\eta$)-plane.
Using the relations above this constraint is given by the
equation
\begin{equation}\label{srhoeta}
S=\frac{2\bar\eta [\bar\rho^2+\bar\eta^2-r^2-\bar\rho(1-r^2)+
       (\bar\rho^2 +\bar\eta^2-1)r \cos\phi]}{((1-\bar\rho)^2+\bar\eta^2)
         (\bar\rho^2+\bar\eta^2+r^2 +2 r\bar\rho \cos\phi)}
\end{equation}
Similarly the relation between $C$ and $\bar\rho$, $\bar\eta$ reads
\begin{equation}\label{crhoeta}
C=\frac{2 r\bar\eta\, \sin\phi}{
   \bar\rho^2+\bar\eta^2+r^2 +2 r\bar\rho \cos\phi}
\end{equation}
This is equivalent to 
\begin{equation}\label{ccirc}
(\bar\rho + r\cos\phi)^2 + \left(\bar\eta -\frac{r\sin\phi}{C}\right)^2
=\left(\frac{1}{C^2}-1\right)(r \sin\phi)^2
\end{equation}
describing a circle in the ($\bar\rho$, $\bar\eta$)-plane
with centre at ($-r\cos\phi$, $(r\sin\phi)/C$) and radius
$r\sin\phi\sqrt{1-C^2}/C$.

The current experimental results for $S$ and $C$ are
\begin{equation}
\begin{array}{cc}
S=+0.02\pm 0.34\pm 0.05 \quad ({\rm BaBar} \cite{BABAR1}) &
-1.00\pm 0.21 \pm 0.07 \quad ({\rm Belle} \cite{BELLE1}) \\
 & \\
C=-0.30\pm 0.25\pm 0.04 \quad ({\rm BaBar} \cite{BABAR1}) &
-0.58\pm 0.15\pm 0.07 \quad ({\rm Belle} \cite{BELLE1}) 
\end{array}
\end{equation}
A recent preliminary update from BaBar gives \cite{HFAG}
\begin{equation}\label{scprel}
S=-0.40\pm 0.22\pm 0.03 \qquad C=-0.19\pm 0.19\pm 0.05
\end{equation}

\section{Penguin Contribution}

In this section we discuss theoretical calculations
of the penguin contribution in $B\to\pi^+\pi^-$.
The analysis is based on the effective weak hamiltonian
\begin{equation}\label{heff}
{\cal H}_{eff}=\frac{G_F}{\sqrt{2}}\sum_{p=u,c}\lambda_p
\left(C_1\, Q^p_1 + C_2\, Q^p_2 +\sum_{i=3,\ldots,6,\, 8g}C_i\, Q_i\right)
+ {\rm h.c.}
\end{equation}
where $C_i$ are Wilson coefficients, known at next-to-leading
order \cite{BJLW}, and $Q_i$
($i=1,\ldots ,6$) are local four-quark operators with flavour 
structure $(\bar db)(\bar qq)$, $q=u,d,s,c,b$. $Q_{8g}$ is the
chromomagnetic operator $\sim m_b\bar d\sigma\cdot G(1+\gamma_5)b$. 
The CKM factors are here denoted by $\lambda_p=V^*_{pd}V_{pb}$.

\subsection{QCD Factorization}

The penguin parameter $r\, e^{i\phi}$ has been computed
in \cite{BBNS3} in the framework of QCD factorization.
The result can be expressed in the form
\begin{equation}\label{rqcd}
r\, e^{i\phi}= -
\frac{a^c_4 + r^\pi_\chi a^c_6 + r_A[b_3+2 b_4]}{
 a_1+a^u_4 + r^\pi_\chi a^u_6 + r_A[b_1+b_3+2 b_4]}
\end{equation}
where we neglected the very small effects from electroweak
penguin operators.
The factorization coefficients $a_i$ are linear combinations
of the Wilson coefficients $C_i$ in the effective weak Hamiltonian
and include the ${\cal O}(\alpha_s)$ corrections from
hard gluon interactions in the weak matrix elements.
Their expressions can be found in \cite{BBNS3}. The quantities
$r^\pi_\chi$ and $r_A$ are defined by
\begin{equation}\label{rchira}
r^\pi_\chi (\mu) = \frac{2 m^2_\pi}{\bar m_b(\mu)(\bar m_u(\mu)+\bar m_d(\mu))}
\qquad r_A=\frac{f_B f_\pi}{m^2_B F^{B\to\pi}_0(0)}
\end{equation}
$r^\pi_\chi$ is defined in terms of the $\overline{MS}$ quark masses
$\bar m_q(\mu)$ and depends on the renormalization scale $\mu$.
$F^{B\to\pi}_0(0)$ is a $B\to\pi$ transition form factor, evaluated at
momentum transfer $q^2=m^2_\pi\simeq 0$.

Both quantities in (\ref{rchira}) are of subleading power
$r^\pi_\chi\sim r_A\sim \Lambda_{QCD}/m_b$.
$r_A\approx 0.003$ is numerically very small. It sets the scale
for the weak annihilation effects in the amplitude, which are parametrized
by the $b_i$ in (\ref{rqcd}) \cite{BBNS3}.  
They represent power corrections that are not calculable in QCD factorization.
Model-dependent estimates for these subleading effects have been given
in \cite{BBNS3} in order to assess the corresponding uncertainties.
On the other hand, $r^\pi_\chi(1.5\,{\rm GeV})\approx 0.7$ is
numerically sizable. Still the important penguin contributions
$a^p_6$, $p=u$, $c$, are calculable and can be included in the analysis.
A third class of power corrections that need to be considered are
uncalculable spectator interactions, some of which also come with the
parameter $r^\pi_\chi$. These enter the $a_i$ in (\ref{rqcd}) and
were also estimated in \cite{BBNS3}.

In Table \ref{tab:rphi} we show the values for $r$ and $\phi$ 
\renewcommand{\arraystretch}{1.3}
\begin{table}[htb]
\begin{center}
\begin{tabular}{|c|c|c|c|c|c|c|c|}\hline\hline
&
$\mu$&
$m_u+m_d$&
$m_c$&
$f_B$&
$F_0^{B\to\pi}$&
$\alpha_2^{\pi}$&
$\lambda_B$
\\ \hline
  $r=0.107$
& $\pm0.005$
& $\pm0.019$
& $\pm0.002$
& $\pm0.003$
& $\pm0.002$
& $\pm0.003$
& $\pm0.002$
\\ \hline
  $\phi=0.150$
& $\pm0.023$
& $\pm0.001$
& $\pm0.057$
& $\pm0.002$
& $\pm0.002$
& $\pm0.003$
& $\pm0.002$
\\ \hline
\end{tabular}

\vspace*{0.3cm}

\begin{tabular}{|c|c|c|}\hline
&
$(\rho_H,\phi_H)$&
$(\rho_A,\phi_A)$
\\ \hline
  $r=0.107$
& $\pm0.001$
& $\pm0.024$
\\ \hline
  $\phi=0.150$
& $\pm0.010$
& $\pm0.24$
\\ \hline\hline
\end{tabular}
\end{center}
\centerline{\parbox{14cm}{\it{
\caption{\label{tab:rphi}
Theoretical values for $r$ and $\phi$ and their
uncertainties from various sources within QCD factorization.
The upper part displays uncertainties from input into the
factorization formulas. The lower part
gives the uncertainties from a model estimate of power corrections
(see text for details). 
}}}}
\end{table}
from a calculation within the QCD factorization framework
as described in \cite{BBNS3}. We also display the uncertainties
from various sources, distinguishing two classes.
In the upper part we give the uncertainties from input into the
factorization formulas at next-to-leading order, as well as the
sensitivity to the renormalization scale $\mu$. This input
is defined in Table \ref{tab:input}.
\renewcommand{\arraystretch}{1.3}
\begin{table}[htb]
\begin{center}
\begin{tabular}{|c|c|c|c|c|c|}\hline\hline
$m_u+m_d$&
$m_c(m_b)$&
$f_B$&
$F_0^{B\to\pi}(0)$&
$\alpha_2^{\pi}$&
$\lambda_B$
\\ \hline
 $0.0091\pm 0.0021$
& $1.3\pm 0.2$
& $0.18\pm 0.04$
& $0.28\pm 0.05$
& $0.1\pm 0.3$
& $0.35\pm 0.15$
\\ \hline\hline
\end{tabular}
\end{center}
\centerline{\parbox{14cm}{\it{
\caption{\label{tab:input}
Input used for Table \ref{tab:rphi}.
We take $\mu\in [m_b/2, 2 m_b]$.
The values for $m_u+m_d\equiv (m_u+m_d)(2 {\rm GeV})$, $m_c(m_b)$,
$f_B$ and $\lambda_B$ are in GeV.  
}}}}
\end{table}
The second class of uncertainty is due to the model
estimates employed for power corrections.
As in \cite{BBNS3} these effects are parameterized
by phenomenological quantities
\begin{equation}\label{xhxa}
X_{H,A}=\left(1+\rho_{H,A}\,e^{i\phi_{H,A}}\right)\ln\frac{m_B}{\Lambda_h}
\end{equation}
that enter power corrections to hard spectator scattering ($H$)
and weak annihilation effects ($A$). The default values have
$\rho_{H,A}=0$. They depend on an infra-red cut-off parameter $\Lambda_h$,
which we take as $\Lambda_h=0.5\,{\rm GeV}$. 
An error of $100\%$ is then assigned to this estimate by allowing
for arbitrary phases $\phi_H$, $\phi_A$ and taking $\rho_H$, $\rho_A$
between $0$ and $1$. 
The impact on $r$ and $\phi$ of this second class of uncertainties
is shown in the lower part of Table \ref{tab:rphi} and is seen to be 
completely dominated by the annihilation contributions.

Adding the errors in quadrature we find
\begin{eqnarray}
r &=& 0.107 \pm 0.020 \pm 0.024\\
\phi &=& 0.15 \pm 0.06 \pm 0.24
\end{eqnarray}
where the first (second) errors are from the first
(second) class of uncertainties. Combining both in quadrature
we finally arrive at 
\begin{equation}\label{rphi}
r=0.107\pm 0.031 \qquad \phi=0.15\pm 0.25
\end{equation}
which we take as our reference predictions for $r$ and $\phi$
in QCD factorization.

\subsection{Expansion in $1/m_b$ and $1/N$}

In order to obtain additional insight into the structure of
hadronic $B$-decay amplitudes, it will be interesting to consider these
quantities in a simultaneous expansion in powers of $1/m_b$ and
$1/N$, where $N$ is the number of colours.
Expanding in $1/m_b$ alone corresponds to the framework of
QCD factorization, implying naive factorization at leading order,
which receives calculable corrections. Large-$N$ expansion of weak
decay amplitudes gives an entirely different justification for
naive factorization, which holds at leading order in $1/N$.
The large-$N$ limit has previously been applied to weak decays
of kaons and $D$ mesons \cite{BG,BGR,BBG}.
Here we would like to explore the consequences of combining the
heavy-quark and large-$N$ limits in the analysis of important
subleading corrections to naive factorization.

For this purpose we treat the Wilson coefficients $C_1$, $C_2$,
$C_4$, $C_6$ and $C_{8g}$ as quantities of order one. 
In the limit $N\to\infty$ strictly speaking only $C_1$ and $C_{8g}$
are nonvanishing. However, $C_2$, $C_4$ and $C_6$ vanish
slower than $1/N$, as $N\to\infty$, if the large logarithm $\ln M_W/m_b$
is considered $\sim N$, in accordance with the usual renormalization
group (RG) counting $\alpha_s\ln M_W/m_b\sim 1$ and with $\alpha_s\sim 1/N$.  
More specifically, $C_2$, $C_4$ $\sim\ln N/N$ and $C_6\sim 1/N^{2/11}$.
The formal treatment of these coefficients as order unity is identical
to the usual counting of the coefficients in RG improved perturbation
theory. From this latter case it is also clear that the small numerical 
size of $C_2$ and especially of the penguin coefficients 
$C_4$, $C_6$ is related to small anomalous dimensions, which are
small accidentally, but not because of a particular parametric suppression.
The coefficients $C_3$, $C_5$ on the other hand are suppressed relative to 
$C_4$ and $C_6$ by an explicit factor of $N$. We will thus take
$C_3$, $C_5\sim 1/N$. Similar considerations can be found in 
\cite{BGR,BBG}.
Concerning the heavy-quark limit $m_b\gg\Lambda_{QCD}$ there is no difference
to the conventional counting in RG improved perturbation theory, 
where $M_W\gg m_b$ is assumed.

After discussing the Wilson coefficients we next turn to the hadronic
matrix elements in QCD factorization. We first need to determine how
the various quantities entering these matrix elements scale for
large $m_b$ and $N$. Based on these results we shall then expand
the factorization coefficients $a_i$ and $b_i$ to next-to-leading
order in a double expansion in $1/m_b$ and $1/N$. 
That is, we keep terms of order one, as well as corrections 
suppressed by either a single power of $1/m_b$ or $1/N$. We neglect terms
that exhibit a suppression by two or more powers of the expansion
parameters $1/m_b$ or $1/N$, such as $1/m^2_b$, $1/N^2$ and $1/m_b N$.
It turns out that in this approximation all subleading contributions
suffering from infrared endpoint singularities in the QCD factorization
approach are absent. This includes both spectator interactions of
subleading twist and all weak annihilation amplitudes. On the other hand,
important effects as hard QCD corrections or the penguin contribution from
$Q_6$, which is formally power suppressed, are retained.

Let us postpone annihilation effects for the moment and examine
first the coefficients $a_i$. The most important ones for our
purpose are $a_1$, $a_4$ and $a_6$, which may be written as \cite{BBNS3}
\begin{eqnarray}
a_1 &=& C_1+\frac{C_2}{N}\left[1+\frac{C_F\alpha_s}{4\pi}V_\pi\right]
           +\frac{C_2}{N}\frac{C_F\pi\alpha_s}{N}H_{\pi\pi} \label{a1}\\ 
a^p_4 &=& C_4+\frac{C_3}{N}\left[1+\frac{C_F\alpha_s}{4\pi}V_\pi\right]
           +\frac{C_F\alpha_s}{4\pi}\frac{P^p_{\pi,2}}{N}
           +\frac{C_3}{N}\frac{C_F\pi\alpha_s}{N}H_{\pi\pi} \label{a4p}\\ 
a^p_6 &=& C_6+\frac{C_5}{N}\left[1-6\frac{C_F\alpha_s}{4\pi}\right]
           +\frac{C_F\alpha_s}{4\pi}\frac{P^p_{\pi,3}}{N} \label{a6p}
\end{eqnarray}
Here $V_\pi$, $P^p_{\pi,2}$, $P^p_{\pi,3}$ are calculable quantities
of order one (in both $1/m_b$ and $1/N$), containing convolution
integrals over pion light-cone distribution amplitudes \cite{BBNS3}.
The coefficient $H_{\pi\pi}$ describes hard spectator scattering
and reads
\begin{eqnarray}\label{hpp}
H_{\pi\pi} &=& \frac{f_B f_\pi}{m^2_B F^{B\to\pi}_0(0)}
\int^1_0\frac{d\xi}{\xi}\phi_B(\xi)\,
\int^1_0\frac{d x}{\bar x}\phi_\pi(x)\,
\int^1_0\frac{d y}{\bar y} \left[
  \phi_\pi(y)+r^\pi_\chi \frac{\bar x}{x}\phi_p(y)\right] \nonumber\\
&\equiv& H_{\pi\pi,2}+H_{\pi\pi,3}
\end{eqnarray}
Here $\phi_B$ is the leading-twist light-cone distribution amplitude
of the $B$ meson, $\phi_\pi$ the one of the pion. 
$\phi_p(y)=1$ is the two-particle, twist-3 component of the pion
light-cone wave function. We recall that the correction
$\sim r^\pi_\chi\phi_p$ (defined as $H_{\pi\pi,3}$ in (\ref{hpp})) 
is uncalculable, as indicated
by the end-point divergence in the $y$-integral, but it is power
suppressed in $1/m_b$.

The large-$m_b$, large-$N$ scaling of the various terms is as follows:
\begin{equation}\label{phibpipxi}
\phi_B, \phi_\pi, \phi_p \sim 1\qquad \xi\sim 1/m_b \qquad x,y\sim 1
\end{equation}
\begin{equation}\label{rffbpi}
r^\pi_\chi\sim 1/m_b\qquad F^{B\to\pi}_0(0)\sim 1/m^{3/2}_b
\qquad f_B\sim N^{1/2}/m^{1/2}_b\qquad f_\pi\sim N^{1/2}
\end{equation}
We then have
\begin{equation}\label{hpp23}
H_{\pi\pi,2}\sim N \qquad H_{\pi\pi,3}\sim N/m_b
\end{equation}
Further we note $C_F\sim N$, $\alpha_s\sim 1/N$.
Expanding (\ref{a1}) - (\ref{a6p}) to first order in $1/m_b$ and
$1/N$ we find
\begin{equation}\label{a1mn}
a_1\,\dot=\, C_1 +\frac{C_2}{N}\left[1+\frac{C_F\alpha_s}{4\pi}V_\pi\right]
           +\frac{C_2}{N}\frac{C_F\pi\alpha_s}{N}H_{\pi\pi,2} 
\end{equation}
\begin{equation}\label{a46mn}
a^p_4\,\dot=\, C_4 +\frac{C_F\alpha_s}{4\pi} \frac{P^p_{\pi,2}}{N}
\qquad\quad
r^\pi_\chi a^p_6\,\dot=\, r^\pi_\chi C_6
\end{equation}
We observe that to this order in the double expansion, the uncalculable
power correction $\sim H_{\pi\pi,3}$ does not appear in $a_1$, to
which it only contributes at order $1/m_b N$. On the other hand, 
the leading-twist effect $\sim H_{\pi\pi,2}$ is retained as well as the 
vertex corrections $\sim V_{\pi\pi}$. Both contribute to $a_1$ at order
$1/N$. For $a^p_4$ the hard-spectator term is absent altogether because
it scales as $1/N^2$, but the nontrivial penguin loop corrections
still contribute at order $1/N$.
Since $r^\pi_\chi a^p_6$ is already $\sim 1/m_b$ we omit all
$1/N$ effects.

We next show that within our approximation the expressions in
(\ref{a1mn}), (\ref{a46mn}) receive no further corrections from
weak annihilation. From \cite{BBNS3} we recall that the annihilation
coefficients $b_i$ appearing in (\ref{rqcd}) can be written as
\begin{equation}\label{b13}
b_1=\frac{C_F}{N^2}C_1 A^i_1\qquad
b_4=\frac{C_F}{N^2}\left[C_4 A^i_1 +C_6 A^i_2 \right]
\end{equation}
\begin{equation}\label{b4}
b_3=\frac{C_F}{N^2}\left[C_3 A^i_1 +C_5(A^i_3+A^f_3)+N C_6 A^f_3\right]
\end{equation}
The parameters $A^{i,f}_k$ are not calculable in QCD factorization.
However, they have been estimated in \cite{BBNS3} from a diagrammatic
analysis of the annihilation topologies in a way that keeps track
of the correct counting in $1/m_b$ and $1/N$. One finds \cite{BBNS3}
\begin{equation}\label{aifra}
A^i_{1,2} \sim 1/N\qquad
A^{i,f}_3 \sim 1/m_b N\qquad\quad r_A \sim N/m_b
\end{equation}
and therefore
\begin{equation}\label{rab124}
r_A b_{1,4} \sim 1/m_b N \qquad\quad
r_A b_3 \sim 1/m^2_b
\end{equation}
All annihilation effects thus contribute to $r$ in (\ref{rqcd})
only at second order in the double expansion, as anticipated above.
Numerically the largest impact on $r$ comes from $b_3$, which in
turn is dominated by the term $\sim C_6$. The contribution to $b_3$
from $C_3$ is highly suppressed $\sim 1/m_b N^2$, the one from $C_5$
even by $1/m^2_b N^2$, and both are very small numerically.
We remark that, unlike all other corrections to naive factorization,
the annihilation term $r_A b_3$ is leading in the large-$N$ limit.

We conclude that (\ref{a1mn}) and (\ref{a46mn}) give indeed the
amplitude coefficients complete through first order in $1/m_b$ and
$1/N$. We stress again that in this approximation these quantities
are fully calculable. In other words, the problematic corrections,
uncalculable in QCD factorization, from higher-twist spectator 
interactions and weak annihilation are at least doubly suppressed
in the combined heavy-quark, large-$N$ expansion.
This observation shows that the large-$N$ limit yields a useful
organizing principle complementary to the $1/m_b$ expansion.
If the large-$N$ limit is not entirely unrealistic, the double
expansion will provide an additional tool to improve our theoretical
control over two-body hadronic $B$-decay amplitudes.
Experience with similar decays of $K$ and $D$ mesons suggests that
considerations based on large-$N$ arguments can be a reasonable
approach to these problems \cite{BG,BGR,BBG}. After all, as discussed in
this paper, for some applications approximate results are sufficient
and precise calculations are not necessarily required.

It is interesting to evaluate the approximations (\ref{a1mn}),
(\ref{a46mn}) numerically and to compare with the full NLO QCD factorization
results that include the estimates of uncalculable power corrections.
Using default input parameters we quote three values
for the various coefficients: The first, second and third numbers
give, respectively, the result for QCD factorization, for the
$1/m_b$, $1/N$ approximation in (\ref{a1mn}), (\ref{a46mn}), and for
naive factorization.\footnote{The coefficients in naive factorization
are $a_i=C_i+C_{i-(-1)^i}/3$, with leading-log values for the $C_i$,
and $b_i=0$.}
\begin{equation}
\begin{array}{rccc}
a_1 = & 1.00+0.02 i & 1.02+0.02i & 1.03    \\
a^c_4 = & -0.033-0.007i & -0.038-0.006i & -0.027    \\
r^\pi_\chi a^c_6 = & -0.056-0.007i & -0.041 & -0.038    \\
r = &  0.107 & 0.084 & 0.068 \\
\phi = & 0.150 & 0.065 & 0
\end{array}
\end{equation}
For comparison, the default value for the annihilation correction
to the penguin amplitude is $r_A(b_3+2 b_4)=-0.010$
in the model of \cite{BBNS3}.
We see that to first order in the double expansion we recover
the QCD factorization result to a large extent.
Since we use the full NLO coefficient $C_4$ in $a^c_4$, there is strictly
speaking an ambiguity whether or not to include in $a^c_4$ the $C_3$ terms, 
which cancel a small part of the NLO scheme dependence.
If we  include $C_3$ in the vertex and penguin correction part
of $a^c_4$, still neglecting hard spectator scattering,
we find $-0.036-0.007i$ instead of the $-0.038-0.006i$ above,
and $(r,\phi)$ becomes $(0.081,0.083)$ instead of $(0.084,0.065)$.
Apparently the $1/N$ approximation proposed here does not
seem to be entirely unrealistic, at least within the standard
QCD factorization framework at next-to-leading order.

\section{Unitarity Triangle from $S$ and $\sin 2\beta$}

\subsection{Determining $\bar\rho$ and $\bar\eta$}

In this section we discuss the determination of the unitariy
triangle by combining the information from $S$
with the value of $\sin 2\beta$, which is known with high 
precision from CP violation measurements in $B\to J/\Psi K_S$.
As we shall see, this method allows for a particularly transparent
analysis of the various uncertainties.
Both $\bar\rho$ and $\bar\eta$ can be obtained, which fixes the
unitarity triangle. A comparison with other determinations then
provides us with a test of the standard model.

The angle $\beta$ of the unitarity triangle is given by
\begin{equation}\label{taus2b}
\tau\equiv\cot\beta=\frac{\sin 2 \beta}{1-\sqrt{1-\sin^2 2\beta}}
\end{equation}
The current world average \cite{HFAG} 
\begin{equation}\label{sin2bexp}
\sin 2\beta =0.739\pm 0.048
\end{equation}
implies
\begin{equation}\label{tauexp}
\tau=2.26\pm 0.22
\end{equation}
Given a value of $\tau$, $\bar\rho$ is related to $\bar\eta$
by
\begin{equation}\label{rhotaueta}
\bar\rho = 1-\tau\, \bar\eta
\end{equation}
The parameter $\bar\rho$ may thus be eliminated from $S$
in (\ref{srhoeta}), which can be solved for $\bar\eta$ to yield
\begin{equation}\label{etataus}
\bar\eta=\frac{(1+\tau S)(1+r \cos\phi)-\sqrt{(1-S^2)(1+r^2+2 r\cos\phi)-
  (1+\tau S)^2 r^2 \sin^2\phi}}{(1+\tau^2)S}
\end{equation}

So far, no approximations have been made and
eqs. (\ref{rhotaueta}), (\ref{etataus}) are still completely general.
The two observables $\tau$ (or $\sin 2\beta$) and $S$ determine
$\bar\eta$ and $\bar\rho$ once the theoretical penguin parameters
$r$ and $\phi$ are provided. It is at this point that some theoretical
input is necessary. We will now consider the impact of the
parameters $r$ and $\phi$, and of their uncertainties, on the analysis.

We first would like to point out that the sensitivity of $\bar\eta$
in (\ref{etataus}) on the strong phase $\phi$  is rather mild. In fact,
the dependence on $\phi$ enters in (\ref{etataus}) only at second order.
Expanding in $\phi$ we obtain to lowest order
\begin{equation}\label{etataus0}
\bar\eta\dot=\frac{1+\tau S -\sqrt{1-S^2}}{(1+\tau^2)S}(1+r)
\end{equation}
This result is corrected at second order in $\phi$ through
\begin{equation}\label{deletatau}
\Delta\bar\eta=\left(\frac{1-S^2+r(1+\tau S)^2}{(1+r)\sqrt{1-S^2}}-(1+\tau S)
\right)\,\frac{r\phi^2}{2(1+\tau^2)S}
\end{equation}
This feature is very welcome since it is in particular the
strong phase that is difficult to calculate with good precision.
Nevertheless, we know from factorization in the heavy-quark limit
that the strong phase is suppressed either by $\alpha_s$, if it arises
from hard scattering, or by $\Lambda_{QCD}/m_b$ for soft corrections.
This means that even if $\phi$ is not accurately known, it will
have very little impact on $\bar\eta$ as long as it is of moderate
size. Since $r$ is also small, the second order effect from $\phi$
in (\ref{deletatau}) is even further reduced.
As an example, for $S=0$ one finds
\begin{equation}\label{deleta0}
\Delta\bar\eta=-\frac{1-r}{1+r}\frac{\tau}{1+\tau^2}\frac{r}{2}\phi^2
\end{equation}
For typical values $r\approx 0.1$, this implies that 
$|\Delta\bar\eta| < 0.01$ for $\phi$ up to $45^\circ$, which is already
a large phase. For $\phi < 20^\circ$, which is more realistic,
one has a negligible shift $|\Delta\bar\eta| < 0.002$.
Consequently, the relation in (\ref{etataus0}) is most likely
a very good approximation to the exact result.
Note that apart from neglecting the phase $\phi$, no approximation
is made in (\ref{etataus0}). The resulting expression
is strikingly simple. $\bar\eta$ is essentially determined 
by the CP violating observables $S$ and $\tau$. The only dependence  
on the penguin parameter $r$ is through an overall factor 
of $(1+r)$. Typically $r\approx 0.1$, as predicted in QCD factorization
but indicated also by other approaches. The effect is again a
fairly small correction. A $100\%$ uncertainty on this estimate
of $r$ would translate into a $10\%$ uncertainty in $\bar\eta$.

\begin{figure}[t]
\psfrag{S}{$S$}
\psfrag{etabar}{$\bar\eta$}
\begin{center}
\psfig{figure=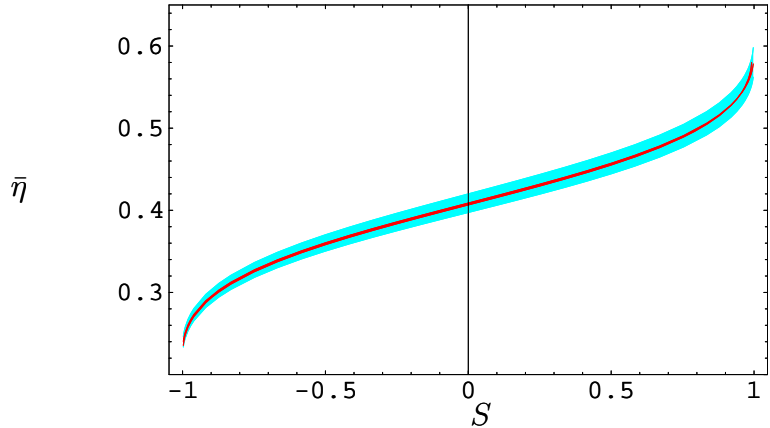}
\end{center}
{\it{\caption{CKM phase $\bar\eta$ as a function of the mixing-induced
CP asymmetry $S$ in $B\to\pi^+\pi^-$ within the standard model
for $\sin 2\beta=0.739$.
The dark (light) band reflects the theoretical uncertainty
in the penguin phase $\phi=0.15\pm 0.25$ 
(penguin amplitude $r=0.107\pm 0.031$).      
\label{fig:etabspp}}}}
\end{figure}

The determination of $\bar\eta$ as a function of $S$ is shown
in Fig. \ref{fig:etabspp},
which displays the theoretical uncertainty from the penguin
parameters $r$ and $\phi$ in QCD factorization.

In the determination of $\bar\eta$ and $\bar\rho$ described
here discrete ambiguities do in principle arise.
One source is the well-known ambiguity in relating $\sin 2\beta$
to a value of $\beta$, or equivalently $\tau=\cot\beta$.
Apart from the solution shown in (\ref{taus2b}), a second
solution exists with the sign of the square root reversed.
It corresponds to a larger value of $\beta$, incompatible
with the standard fit of the unitarity triangle.
An additional ambiguity comes from the second solution
for $\bar\eta$, which is the result given in (\ref{etataus}) with a
positive sign in front of the square root.
This case may be considered separately, but will usually also yield
solutions in conflict with other information on the CKM phases.

\subsection{Standard Model Prediction for $S$}

In this section we shall use theoretical information on
$r$ and $\phi$ based on QCD factorization to compute the value
of $S$ expected within the standard model.
Using (\ref{rhotaueta}), one can write (\ref{srhoeta}) in the
form
\begin{equation}\label{staueta}
S=2\,\frac{\frac{1+\tau^2}{1+r\cos\phi}\,\bar\eta-\tau\left(
          1+\frac{r^2\sin^2\phi}{(1+r\cos\phi)^2}\right)}{
   1+\left(\frac{1+\tau^2}{1+r\cos\phi}\,\bar\eta-\tau\right)^2 +
    (1+\tau^2)\frac{r^2\sin^2\phi}{(1+r\cos\phi)^2}}
\end{equation}
Since the terms $\sim r^2\sin^2\phi\, (< 10^{-3})$ are very small,
$S$ is well approximated by
\begin{equation}\label{szaprox}
S=\frac{2z}{1+z^2}\qquad\quad 
 z\equiv\frac{1+\tau^2}{1+r\cos\phi}\,\bar\eta-\tau
\end{equation}
Taking \cite{CKMF}
\begin{equation}\label{tauetaexp}
\tau=2.26\pm 0.22\qquad \bar\eta=0.35\pm 0.04
\end{equation}
and 
\begin{equation}\label{rphinum}
r=0.107\pm 0.031 \qquad \phi=0.15\pm 0.25
\end{equation}
we find from (\ref{staueta})
\begin{equation}\label{spred}
S=-0.59\quad ^{+0.18}_{-0.11}\,(\tau)\quad ^{+0.38}_{-0.25}\,(\bar\eta)
       \quad ^{-0.07}_{+0.08}\, (r)\quad ^{+0.02}_{-0.00}\,(\phi)
\end{equation}
We note that the uncertainty from the hadronic phase $\phi$ is negligible
and the uncertainty from the penguin parameter $r$ is rather moderate.
The error from $\tau$ or $\sin 2\beta$, which reflects the current
experimental accuracy in this quantity, is considerably larger. 
The dominant uncertainty, however, is due to $\bar\eta$, which for the
purpose of predicting $S$ has here been taken from a standard CKM fit.
Clearly, the large sensitivity of $S$ to $\bar\eta$ is equivalent
to the fact that in turn $\bar\eta$ has only a fairly weak dependence
on $S$. This feature was already discussed in \cite{BB96}.
Eq. (\ref{spred}) shows that the standard model prefers negative values 
for $S$, but it is difficult to obtain an accurate prediction.

\subsection{Rigorous CKM Bounds with Minimal Hadronic Input}

We will now relax the constraints on the penguin parameters $r$ and $\phi$
coming from direct theoretical calculations and derive rigorous bounds
on $\bar\rho$ and $\bar\eta$ without relying on any detailed information
about hadronic quantities. Specifically, we shall only assume that
the strong phase $\phi$ fulfills
\begin{equation}\label{phi90}
-\frac{\pi}{2}\leq\phi\leq\frac{\pi}{2}
\end{equation}
In view of the fact that $\phi$ is systematically suppressed in
the heavy-quark limit, and that typically $\phi\approx 0.2$
from QCD factorization, this assumption is very weak.

As has been shown in \cite{BS}, the following inequality
can be derived from (\ref{etataus}) for $-\sin 2\beta\leq S\leq 1$
\begin{equation}\label{etb3}
\bar\eta\geq\frac{1+\tau S-\sqrt{1-S^2}}{(1+\tau^2)S}(1+r\cos\phi)
\end{equation}
This bound is still {\it exact} and requires no information
on the phase $\phi$. (The only condition is that $(1+r\cos\phi)$ is
positive, which is no restriction in practice.)

Assuming now (\ref{phi90}), we have $1+r\cos\phi \geq 1$ and
\begin{equation}\label{etabound}
\bar\eta\geq\frac{1+\tau S-\sqrt{1-S^2}}{(1+\tau^2)S}
\quad {\rm if}\quad -\sin 2\beta\leq S\leq 1
\end{equation}
We emphasize that this lower bound on $\bar\eta$ depends only on the
observables $\tau$ and $S$ and is essentially free of hadronic
uncertainties. It holds in the standard model and it is effective under
the condition that $S$ will eventually  be measured in the interval
$[-\sin 2\beta,1]$. Since both $r$ and $\phi$ are expected to be
quite small, we anticipate that the lower limit (\ref{etabound})
is a fairly strong bound, close to the actual value of $\bar\eta$ itself
(compare  (\ref{etataus0})).
We also note that the lower bound (\ref{etabound}) represents the
solution for the unitarity triangle in the limit of vanishing
penguin amplitude, $r=0$. In other words, the model-independent
bounds for $\bar\eta$ and $\bar\rho$ are simply obtained
by ignoring penguins and taking $S\equiv\sin 2\alpha$ when fixing
the unitarity triangle from $S$ and $\sin 2\beta$.

\begin{figure}[p]
\psfrag{S}{$S$}
\psfrag{etab}{$\bar\eta$}
\begin{center}
\psfig{figure=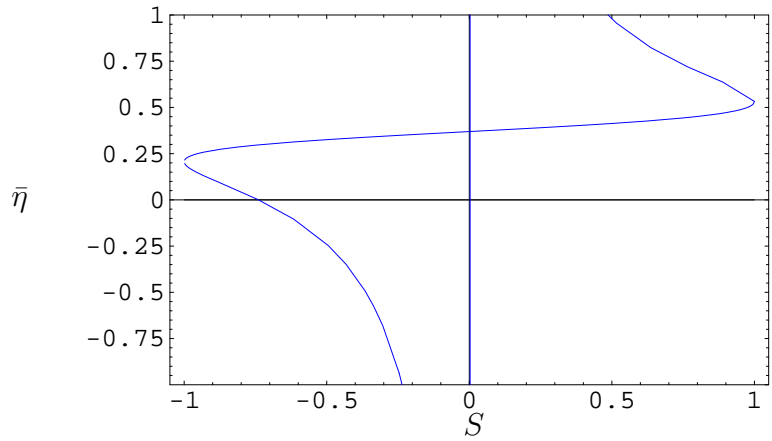}
\end{center}
{\it{\caption{Discrete ambiguities for $\bar\eta$ as a function
of $S$ with $\sin 2\beta=0.739$, $r=0$. For $-\sin 2\beta\leq S\leq 1$
the middle branch defines the lower bound on $\bar\eta$, which is
not affected by the additional solution. \label{fig:etada}}}}
\end{figure}
\begin{figure}[p]
\psfrag{S}{$S$}
\psfrag{etab}{$\bar\eta$}
\begin{center}
\psfig{figure=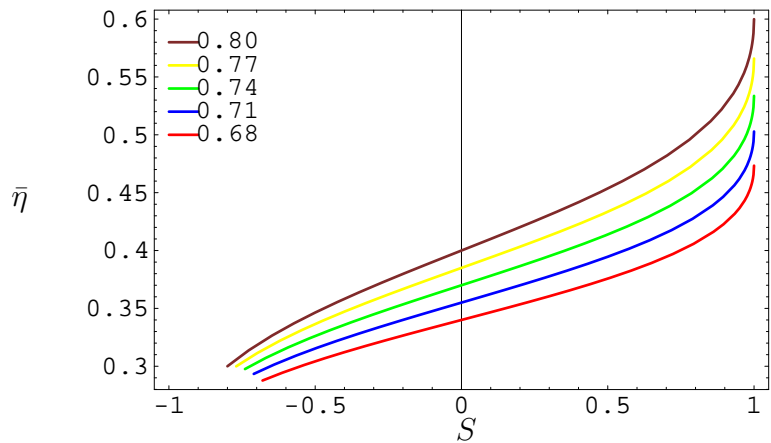}
\end{center}
{\it{\caption{Lower bound on $\bar\eta$ as a function
of $S$ for various values of $\sin 2\beta$
(increasing from bottom to top).  \label{fig:etabound}}}}
\end{figure}
\begin{figure}[p]
\psfrag{S}{$S$}
\psfrag{gamma}{$\gamma$ [deg]}
\begin{center}
\psfig{figure=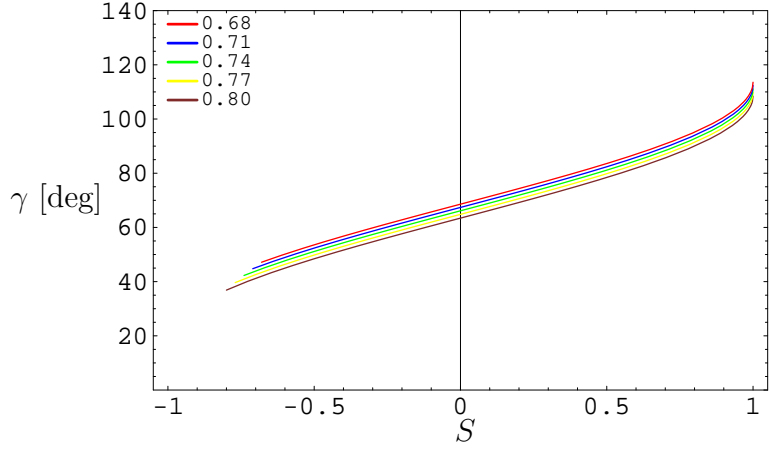}
\end{center}
{\it{\caption{Lower bound on $\gamma$ as a function
of $S$ for various values of $\sin 2\beta$
(decreasing from bottom to top).  \label{fig:gambound}}}}
\end{figure}
\begin{figure}[p]
\psfrag{etab}{$\bar\eta$}
\psfrag{rob}{$\bar\rho$}
\begin{center}
\epsfig{file=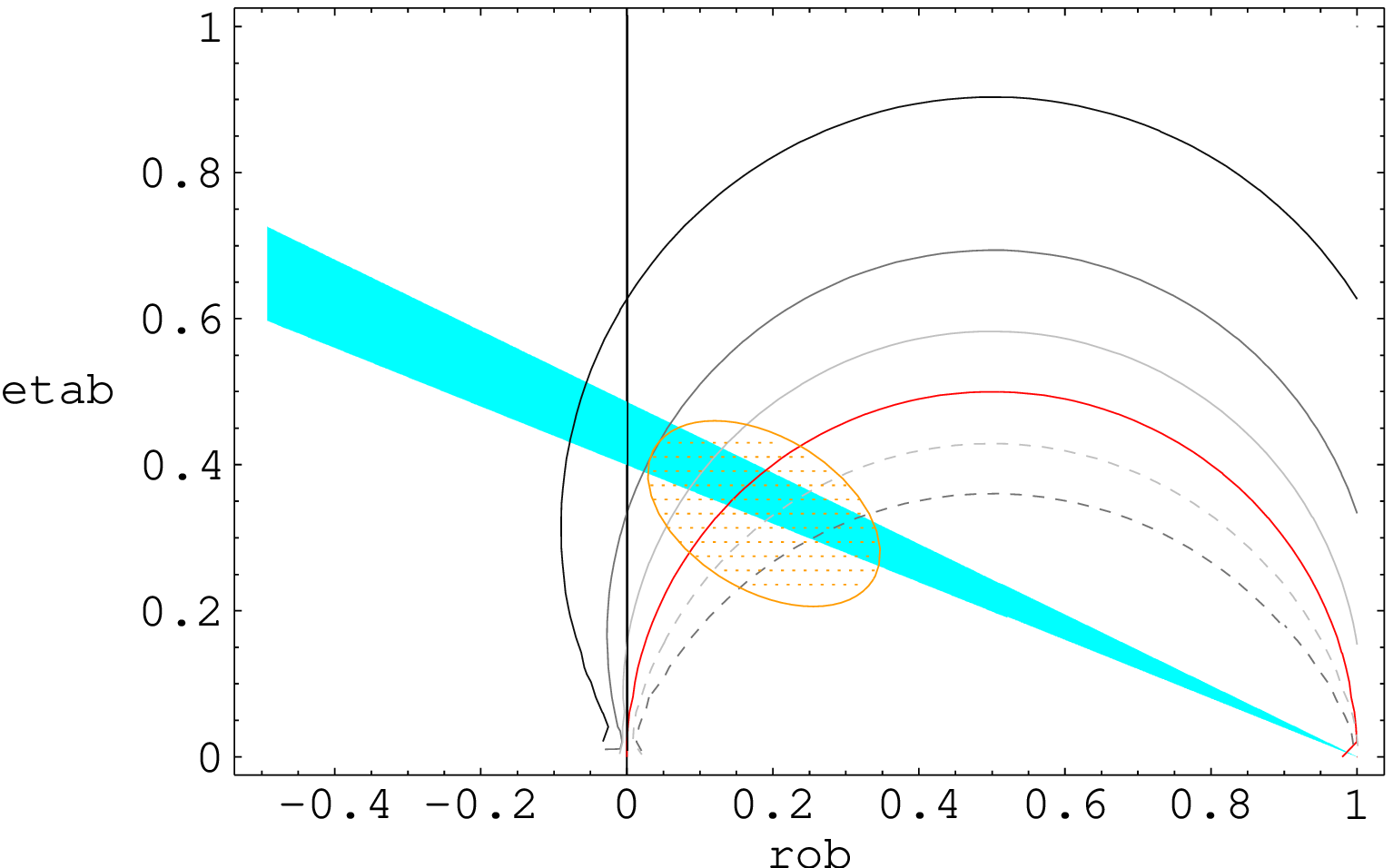,width=14cm,height=8.7cm}
\end{center}
{\it{\caption{Region in the $(\bar\rho,\bar\eta)$ plane
constrained in a model-independent way by $\sin 2\beta=0.739\pm 0.048$
(shaded sector) 
and various possible values for $S$. The allowed area is the part
of the shaded sector to the left of a given line defined by $S$.
These lines correspond, from bottom to top, to $S$=$-0.6$,
$-0.3$, $0$, $0.3$, $0.6$ and $0.9$.
The bound becomes stronger with increasing $S$. 
The result of a standard unitarity triangle fit (dotted ellipse,
from \cite{CKMF}) is overlaid for comparison.   
\label{fig:utbound}}}}
\end{figure}

Let us briefly comment on the second solution for $\bar\eta$, which has 
the minus sign in front of the square root in (\ref{etataus}) replaced by a
plus sign. For positive $S$ this solution is always larger than
(\ref{etataus}) and the bound (\ref{etabound}) is unaffected.
For $-\sin 2\beta\leq S\leq 0$ the second solution gives a negative
$\bar\eta$, which is excluded by independent information on the
unitarity triangle (for instance from indirect CP violation in
neutral kaons ($\varepsilon_K$)). The additional solution for $\bar\eta$
is illustrated in Fig. \ref{fig:etada}
for $r=0$, the case relevant for the lower bound.

Because we have fixed the angle $\beta$, or $\tau$, the lower bound
on $\bar\eta$ is equivalent to an upper bound on $\bar\rho=1-\tau\bar\eta$.
The constraint (\ref{etabound}) may also be expressed as a lower bound
on the angle $\gamma$
\begin{equation}\label{gambound}
\gamma\geq \frac{\pi}{2}-
  \arctan\frac{S-\tau(1-\sqrt{1-S^2})}{\tau S+ 1-\sqrt{1-S^2}} 
\end{equation}
or a lower bound on $R_t$
\begin{equation}\label{rtbound}
R_t\equiv\sqrt{(1-\bar\rho)^2+\bar\eta^2}\geq
\frac{1+\tau S-\sqrt{1-S^2}}{\sqrt{1+\tau^2}S}
\end{equation}
In Figs. \ref{fig:etabound} 
and \ref{fig:gambound}
we represent the lower bound on
$\bar\eta$ and $\gamma$ as a function of $S$ for various values of
$\sin 2\beta$. From Fig. \ref{fig:etabound} we observe that 
the lower bound on $\bar\eta$ becomes stronger as either
$S$ or  $\sin 2\beta$ increase. The sensitivity to $\sin 2\beta$
is less pronounced for the bound on $\gamma$.
Similarly to $\bar\eta$, the minimum allowed value for $\gamma$
increases with $S$. A lower limit $\gamma=90^\circ$ is reached
for $S=\sin 2\beta$. 

In Fig. \ref{fig:utbound} we illustrate the region in the
$(\bar\rho,\bar\eta)$ plane that can be constrained by the
measurement of $\sin 2\beta$ and $S$ using the bound
in (\ref{etabound}).

We finally note that the condition $r\cos\phi > 0$,
which is crucial for the bound, could be independently checked
\cite{SKK} by measuring the mixing-induced CP-asym\-me\-try in
$B_s\to K^+ K^-$. This is because the hadronic physics of
$B_s\to K^+ K^-$ is related to $B_d\to\pi^+\pi^-$ by U-spin
symmetry, a feature that has already been employed
for CKM phenomenology \cite{DF}. 
Our purpose here is to use information from $B_s\to K^+ K^-$
in order to obtain additional input for $B_d\to\pi^+\pi^-$
within the approach suggested above.
To this end we write the CP-violation observable $S$, defined in analogy
to (\ref{acppipi}), for the case of $B_s\to K^+ K^-$.
Neglecting the small phase of $B_s$ -$\bar B_s$ mixing, we find  
\begin{equation}\label{skk}
S(B_s\to K^+ K^-)=\frac{2\bar\eta\, (k r\cos\phi -\bar\rho)}{
 (k r\cos\phi -\bar\rho)^2+\bar\eta^2+k^2 r^2\sin^2\phi}
\end{equation}
where
\begin{equation}\label{kdef}
k\equiv\frac{1-\lambda^2}{\lambda^2}\approx 20
\end{equation}
for Wolfenstein parameter $\lambda=0.22$. Because of the different
CKM hierarchy of the $b\to s$ transition, the penguin contribution
$\sim r$ is strongly enhanced by a factor of $k\approx 20$ compared to
the case of $B_d\to\pi^+\pi^-$. On the other hand, the purely hadronic
quantities $r$ and $\phi$ are identical to the corresponding parameters
for $B_d\to\pi^+\pi^-$ in the limit of exact U-spin symmetry.
We notice that the presumably largest effects from U-spin breaking,
coming from the difference of decay constants $f_K$, $f_\pi$
and form factors for $B_s\to K$ and $B_d\to\pi$ transitions,
largely cancel in the ratio $r$ of penguin over tree amplitudes.
Explorations of further sources of U-spin breaking can be found in
\cite{SKK,MB}. We shall assume that $r\approx 0.1$ as indicated by
QCD factorization. Then the relevant penguin parameter in 
(\ref{skk}) is $k r\approx 2$, which dominates the much smaller
values for $\bar\rho\sim 0.15$. 
As a consequence, (\ref{skk}) predicts, in the standard model,
the sign of $S(B_s\to K^+K^-)$ in correspondence with the sign of
$r \cos\phi$. In QCD factorization this sign is positive and one
expects
\begin{equation}
S(B_s\to K^+K^-)\approx\bar\eta
\end{equation}
A future measurement of this observable will then provide a test
of the assumption made to obtain the above bounds. 
We remark that also from other charmless hadronic $B$ decays
a sizable penguin amplitude is required, independently of
detailed QCD calculations, where $r=0.1$ is a typical value. It is thus
basically excluded that $k r$ will be much below about 2 and that the
term $-\bar\rho$ in the numerator of (\ref{skk}) will be able to compete
so as to change the conclusion. On the other hand, an extreme value of the
phase $\phi\approx \pi/2$, for instance, could be indicated through
the observation of a very small $S(B_s\to K^+K^-)$, which would typically 
amount to a few percent.  
(In the approximation above one would obtain
$S(B_s\to K^+K^-)\approx -\bar\eta\bar\rho/2$, but the mixing phase
could then no longer be neglected. It would tend to further
reduce the asymmetry.)

In this comparison it is legitimate to assume the validity of the
standard model throughout, as the strategy is to look for new physics
via inconsistencies under this assumption, exploiting a multitude of
experimental results and eliminating hadronic uncertainties.

\section{Unitarity Triangle from $S$, $R_b$ and $\sin 2\phi_d$}

In the previous section we have considered a determination of
the unitarity triangle within the standard model. In particular 
we have assumed that the phase of $B_d$-$\bar B_d$ mixing $\beta$, measured
in the time-dependent CP asymmetry of $B\to J/\Psi K_S$,
is indeed the angle $\beta$ of the CKM unitarity triangle, satisfying
the relation $\tan\beta=\bar\eta/(1-\bar\rho)$.
In the presence of new physics this needs no longer to be the case.
From this perspective a different analysis is of interest.
To be specific we shall assume the plausible scenario where the
new physics contributions modify the phase of $B_d$-$\bar B_d$ mixing $\phi_d$,
whereas the $B$ decay amplitudes remain unchanged.
The CP asymmetry in $B\to J/\Psi K_S$ (\ref{sin2bexp}) must then 
be interpreted as the quantity $\sin 2\phi_d$.
Since we can no longer relate $\sin 2\phi_d$ to $\bar\rho$ and $\bar\eta$,
we should fix it to the experimental value in (\ref{sin2bexp})
when using (\ref{scxi}), where $\beta$ is to be replaced by $\phi_d$.
A similar analysis has already been carried out in \cite{FIM}
(see also \cite{BBNS3}).
This paper gives an interesting discussion of the possible implications
of new physics for the case at hand, including a consideration of
rare decay channels. The required penguin-to-tree ratio is estimated
in \cite{FIM} in a phenomenological way. Here we would like to rephrase
the analysis in terms of the hadronic quantities $r$ and $\phi$,
which are more directly related to theoretical input. We thus obtain
formulas in line with our general approach to isolate CKM parameters
and hadronic physics in a transparent way.

Writing
\begin{equation}\label{rhetgam}
\bar\rho=R_b\cos\gamma\qquad
\bar\eta=R_b\sin\gamma\qquad R_b\equiv\sqrt{\bar\rho^2+\bar\eta^2}
\end{equation}
we have 
\begin{equation}\label{srbgam}
S={\rm Im}\left[e^{-2i\phi_d}
\frac{(R_b\cos\gamma+r\cos\phi-i R_b\sin\gamma)^2+r^2\sin^2\phi}{
(R_b\cos\gamma+r\cos\phi)^2 +R^2_b\sin^2\gamma + r^2\sin^2\phi}\right]
\end{equation}
From this relation, for given values of $r$ and $\phi$,
and using the experimental results for $S$, $\sin 2\phi_d$ and $R_b$,
$\gamma$  can be determined. Under the assumptions specified above
this will be the true angle $\gamma$ of the CKM matrix, independent
of the $B_d$-$\bar B_d$ mixing phase $\phi_d$.
Once $\gamma$ is known, $\bar\rho$ and $\bar\eta$
follow from (\ref{rhetgam}).

\begin{figure}[t]
\psfrag{rob}{$\bar\rho$}
\psfrag{etab}{$\bar\eta$}
\begin{center}
\psfig{figure=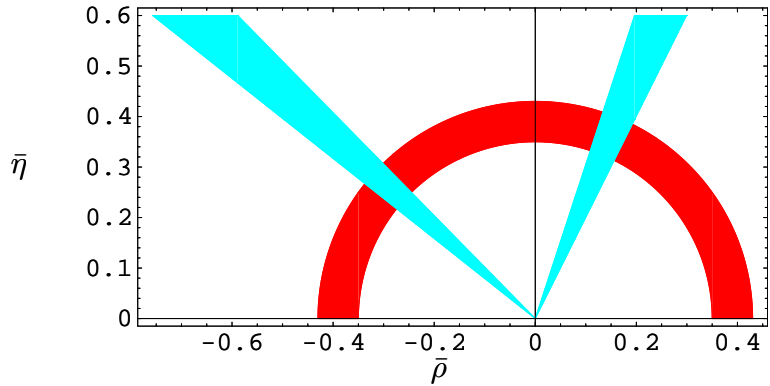}
\end{center}
{\it{\caption{Determination of the unitarity triangle from
$S$, $R_b$ and $\sin 2\phi_d$. We take $\sin 2\phi_d=0.739$,
$R_b=0.39$, the penguin parameters
$r=0.107\pm 0.031$, $\phi=0.15\pm 0.25$, and assume
a hypothetical value of $S=-0.4$.
The resulting ranges of the CKM angle $\gamma$
are $\gamma=67^\circ \pm 4^\circ$ (right band) 
and $\gamma=138^\circ \pm 4^\circ$ (left band)
in this example and reflect the uncertainty in $r$ and $\phi$.
The circular area shows the $R_b$ constraint (\ref{s2pdrbexp}).  
\label{fig:sppnp}}}}
\end{figure}

Experimentally one has \cite{HFAG,CKMF}
\begin{equation}\label{s2pdrbexp}
\sin 2\phi_d=0.739\pm 0.048\qquad\quad R_b=0.39\pm 0.04
\end{equation}
As emphasized in \cite{FIM}, there is a discrete ambiguity
in the sign of $\cos 2\phi_d$, which yields two different
solutions for $\gamma$. The larger value of $\gamma$ will be 
obtained for negative $\cos 2\phi_d$.
The analysis is represented in Fig. \ref{fig:sppnp}, assuming
a particular scenario for illustration and displaying the
impact of the theoretical uncertainty in $r$ and $\phi$.

\section{Direct CP Violation}

So far we have considered the implications of mixing-induced
CP violation, described by $S$.
In the following we shall investigate how useful additional
information can be extracted from a measurement of the direct CP violation
parameter $C$. An alternative discussion of this question
can be found in \cite{GR1}. 

The observable $C$ (see (\ref{crhoeta})) is an odd function of $\phi$.
It is therefore sufficient to restrict the discussion to positive values
of $\phi$. A positive phase $\phi$ is obtained by the perturbative
estimate in QCD factorization, neglecting soft phases with power
suppression. For positive $\phi$ also $C$ will be positive,
assuming $\bar\eta > 0$, and a sign change in $\phi$ will simply
flip the sign of $C$.

In contrast to the case of $S$, the hadronic quantities $r$ and $\phi$
play a prominent role for $C$, as can be seen in (\ref{crhoeta}).
This will in general complicate the interpretation of an experimental
result for $C$.
One aspect of this can be seen as follows. It is usually
expected that small values of the weak phase $\bar\eta$ and 
the strong phase $\phi$ correspond to a small CP asymmetry $C$.
However, in principle this need not be the case. As a counterexample
let us consider the scenario where $\bar\rho=-0.35$, $\bar\eta=0.07$,
$r=0.35$ and $\phi=0.2\approx 11^\circ$. Of course this implies
a very large angle $\gamma$, but this would be possible if the presence of
new physics invalidates the standard unitarity triangle analysis
(still the constraint from $R_b$ is obeyed). Although both $\bar\eta$
and the strong phase $\phi$ are very small, these numbers give
$C=0.995$. More generally, such a situation occurs if
$\bar\rho=-r$, leading to a cancellation in the denominator of $C$.
In this case, assuming also that $\phi$ is small, we get
\begin{equation}\label{cspecial}
C\approx\frac{2\bar\eta r \phi}{\bar\eta^2 + (r\phi)^2}
\end{equation}
which takes its maximal value $C=1$ for $\bar\eta=r\phi$.
Clearly, a scenario of this type requires a very peculiar
coincidence and may seem unlikely. Nevertheless the example
illustrates that the proper interpretation of $C$ can be
rather involved.

The analysis of $C$ becomes more transparent if we fix the
weak parameters and study the impact of $r$ and $\phi$.
An important application is a test of the standard model,
obtained by taking $\bar\rho$ and $\bar\eta$ from a
standard model fit and comparing the experimental result for $C$
with the theoretical expression as a function of $r$ and $\phi$.

Let us first derive a few general results.
An important question is the maximum value of $C$, for given
$\bar\rho$ and $\bar\eta$, allowing an arbitrary variation of 
$r$ and $\phi$.

Varying $r$ we find that $C$ takes its maximum for
\begin{equation}\label{rcmax}
r=R_b\equiv\sqrt{\bar\rho^2 + \bar\eta^2}
\end{equation}
independently of $\phi$.
The resulting maximum $C_{max}(\phi)$ at $r=R_b$ can be written as
\begin{equation}
C_{max}(\phi) =\frac{\sin\gamma\, \sin\phi}{1+\cos\gamma\, \cos\phi}
\end{equation}
and only depends on $\phi$ and $\gamma$. Viewed as a function of $\phi$
it can reach its absolute maximum $C=1$ for $\cos\phi=-\cos\gamma$.

\begin{figure}[t]
\psfrag{phi}{\hspace*{-1cm} $\phi$ [rad]}
\psfrag{r}{$r$}
\begin{center}
\psfig{figure=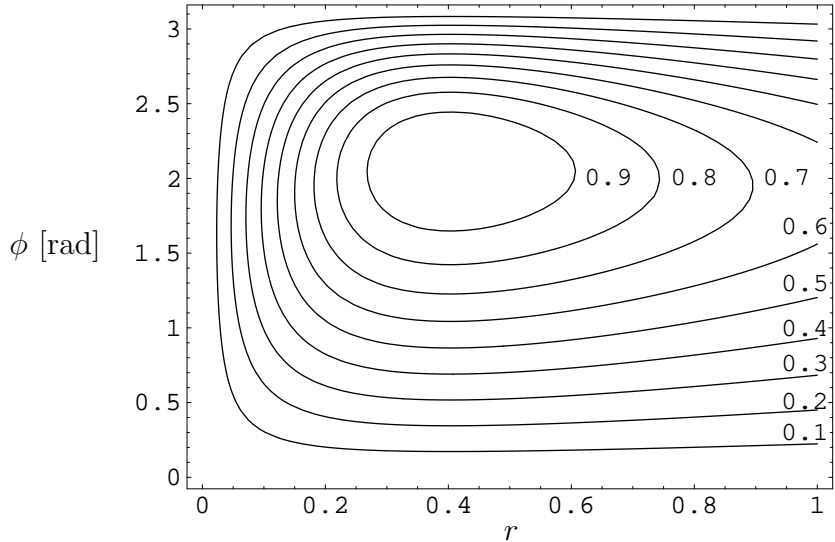}
\end{center}
{\it{\caption{Contours of constant $C$ in the ($r$, $\phi$)-plane
for fixed $\bar\rho=0.20$ and $\bar\eta=0.35$.  \label{fig:cpipi}}}}
\end{figure}

A useful representation is obtained by plotting contours of constant
$C$ in the ($r$, $\phi$)-plane, for given values of $\bar\rho$ and $\bar\eta$.
This is illustrated in Fig. \ref{fig:cpipi}
for the standard model best-fit result $\bar\rho=0.20$, $\bar\eta=0.35$
\cite{CKMF}.
Within the standard model this illustrates the correlations between 
the hadronic penguin parameters $r$ and $\phi$ and direct CP violation
in $B\to\pi^+\pi^-$ decays. Upper limits on $r$ and $\phi$ imply an
upper limit on $C$ unless they acquire unreasonably large values.
For example, $r< 0.15$ and $\phi < 0.5$ yield $C < 0.21$.

We may relax the assumption of the validity of the standard model
and discuss the parameter $C$ from a different perspective.
Similarly to the previous section, we consider the rather general scenario
where new physics renders the standard unitarity triangle fit
to determine $\gamma$ invalid, while the extraction of $R_b$ 
and the $B\to\pi^+\pi^-$ amplitudes remain essentially unaffected.
In this situation it is convenient to slightly rewrite (\ref{crhoeta})
as
\begin{equation}\label{csgamma}
C=\frac{2 \kappa\sin\gamma\, \sin\phi}{
   1+\kappa^2 + 2 \kappa\cos\gamma\, \cos\phi}
\end{equation}
where we have introduced $\kappa\equiv r/R_b=|P/T|$.
If we treat $\gamma$ as unconstrained, we can still place an 
upper bound on $C$ by maximizing $C$ with respect to $\gamma$.
Denoting this maximum by $\bar C$ we find
\begin{equation}\label{barc}
\bar C=\frac{2 \kappa\, \sin\phi}{
  \sqrt{(1+\kappa^2)^2 -4 \kappa^2 \, \cos^2\phi}}
\end{equation}
where the maximum occurs at $\cos\gamma=-2\kappa\cos\phi/(1+\kappa^2)$.

If $\kappa=1$, or equivalently $r=R_b$, then $\bar C\equiv 1$ independent
of $\phi$, and no useful upper bound is obtained.
On the other hand, if $\kappa < 1$, then $\bar C$ is maximized for
$\phi=\pi/2$. Under the general assumptions stated above and
without any assumption on the strong phase $\phi$ we thus find
the general bound
\begin{equation}\label{cbound0}
C < \frac{2\kappa}{1+\kappa^2}
\end{equation}
For the conservative bound $r < 0.15$, $\kappa < 0.38$ this implies
$C < 0.66$. The bound on $C$ can be strengthened by using information
on $\phi$, as well as on $\kappa$, and employing (\ref{barc}).
Then $\kappa < 0.38$ and $\phi < 0.5$ gives $C < 0.39$.

\section{Tests of Factorization Predictions}

The analyses described above require theoretical input on the
penguin parameter $r\exp(i\phi)$. We have relied on 
model-independent calculations
based on the heavy-quark limit of QCD, including model estimates of
subleading effects. To reinforce the validity of the approximations,
in particular the large-$m_b$ limit for realistic values of the
$b$-quark mass, it is important to test other predictions, obtained
within the same framework, against experiment. For this purpose
it is necessary to keep in mind that both hadronic effects as well as
effects from new physics could in general be the origin of any
discrepancy. Both effects need to be disentangled as far as possible.
Especially useful tests of the QCD aspects in hadronic $B$ decays are 
those that have no, or very little, dependence on weak phases and potential
new physics contributions.
We shall discuss several such tests, which pertain to the essential
ingredients for the $P/T$ ratio $r\exp(i\phi)$, namely the tree amplitude,
the penguin amplitude
and annihilation effects in $B\to\pi\pi$ decays.

\subsection{Tree amplitude}

The first example is a factorization test for
the rate of the tree-type decay $B^\pm\to\pi^\pm\pi^0$ by taking
a suitable ratio with the semileptonic rate of $B\to\pi l\nu$.
For similar decays such observables have already been discussed e.g.
in \cite{NS}. More recently it has been proposed to employ
the decay $B\to\pi l\nu$ for an estimate of the tree amplitude
in $B\to\pi^+\pi^-$, assuming factorization \cite{LR}.
This analysis is similar to the one suggested here,
however our main emphasis is somewhat different. 
We consider the factorization test as a cross-check of
dynamical calculations based on the heavy-quark limit \cite{BBNS3},
rather than a way to determine the tree amplitude. The reason for this
distinction is the fact that the tree contribution in $B^+\to\pi^+\pi^0$
is not exactly the same as the one required for $B\to\pi^+\pi^-$. 
A related discussion can also be found in \cite{BN}.

The differential decay rate for $B_d\to\pi^- l^+\nu$
is given by
\begin{equation}\label{bpilnu}
\frac{d\Gamma(B_d\to\pi^- l^+\nu)}{ds}=
\frac{G^2_F m^5_B}{192\pi^3}|V_{ub}|^2\, \lambda^{3/2}_\pi(s)\, f^2_+(q^2)
\end{equation}
with
\begin{equation}\label{larps}
\lambda_\pi(s)=1+r^2_\pi+s^2-2s-2r_\pi-2 r_\pi s
\qquad\quad r_\pi=\frac{m^2_\pi}{m^2_B}\qquad s=\frac{q^2}{m^2_B}
\end{equation}
 Here $q^2$ is the invariant mass of the lepton pair and
$f_+(q^2)$ is a $B\to\pi$ transition form factor. 
Eq. (\ref{bpilnu}) is valid for leptons $l=e$, $\mu$ where the mass
is negligible. The pion mass effect is also very small,
$r_\pi=7\times 10^{-4}$, and can likewise be neglected.
In this case the branching fractions of $B^+\to\pi^+\pi^0$ and
$B_d\to\pi^- l^+\nu$ are related through
\begin{equation}\label{pp0sl}
B(B^+\to\pi^+\pi^0)=3\pi^2 f^2_\pi |V_{ud}|^2
\,\left. \frac{dB(B_d\to\pi^- l^+\nu)}{dq^2}\right|_{q^2=0}\,
\frac{\tau(B^+)}{\tau(B_d)}\, |a_1+a_2|^2
\end{equation}
where $a_1$, $a_2$ are QCD coefficients \cite{BBNS1,BBNS3}.
In (\ref{pp0sl}) the dependence on $|V_{ub}|$ and $f_+(0)$,
which are not known precisely, have cancelled out.
Once the branching fractions are measured, $|a_1+a_2|$ can be
extracted experimentally via (\ref{pp0sl}) and compared
with the theoretical prediction. This test is useful since
neither QCD penguins nor weak annihilation corrections
affect the $B^+\to\pi^+\pi^0$ amplitude. It can thus provide us
with a check on the tree amplitude $a_1+a_2$, which includes
non-trivial hard-spectator interactions in QCD factorization, but does
not depend on the other complications. Moreover,
this test is independent of CKM phases and very unlikely to be 
modified by non-standard physics. It probes, as desired, crucial aspects of
the QCD dynamics in $B\to\pi\pi$ decays. 
  
In order to determine the differential 
semileptonic branching ratio
$dB_{SL}/dq^2$ $\equiv$ $dB(B_d\to\pi^- l^+\nu)/dq^2$ at $q^2=0$,
one needs to fit the $q^2$ spectrum of the semileptonic decay.
We may use the expression 
\begin{equation}\label{dbsl} 
\frac{dB_{SL}}{dq^2}=\left. \frac{dB_{SL}}{dq^2}\right|_{q^2=0}
\,\frac{\left(1-\frac{q^2}{m^2_B}\right)^3
        \left(1+a\frac{q^2}{m^2_{B^*}}\right)^2}{
        \left(1-\frac{q^2}{m^2_{B^*}}\right)^2}
\end{equation}
which follows from (\ref{bpilnu}) and the parameterization 
of the form factor suggested in \cite{LR}
\begin{equation}\label{ffpar} 
\frac{f_+(q^2)}{f_+(0)}=
\frac{1+a\frac{q^2}{m^2_{B^*}}}{1-\frac{q^2}{m^2_{B^*}}}
\end{equation}
At present the data on $B_d\to\pi^- l^+\nu$ decays are not yet accurate 
enough to give a stringent test \cite{LR}. 
The situation should improve substantially in the future and will then
yield valuable information on QCD dynamics in $B\to\pi\pi$ decays.

\subsection{Penguin-to-Tree Ratio}

The decay mode $B^+\to\pi^+ K^0$ is essentially a pure
penguin process, up to a negligible rescattering contribution \cite{BBNS3}.
The ratio of the $B^+\to\pi^+ K^0$ to the $B^+\to\pi^+ \pi^0$
branching fraction is therefore a useful probe of the 
pegnuin-to-tree ratio \cite{BBNS3,BN}.
In analogy to the relevant parameter $r$ in $B\to\pi^+\pi^-$, one
may define a quantity $\tilde r$, which can be expressed through
observables:
\begin{equation}\label{rtilde}
\tilde r\equiv
\left|\frac{(a^c_4+r_\chi a^c_6+r_A b_3)_{\pi K}}{a_1+a_2}\right|
=\left|\frac{V_{ub}}{V_{cb}}\right|\frac{f_\pi}{f_K}
\sqrt{\frac{B(B^+\to\pi^+ K^0)}{2\, B(B^+\to\pi^+\pi^0)}}
=0.099\pm 0.011\pm 0.008
\end{equation}
Here CP-averaged branching fractions are understood. The quoted number
is derived from current experimenatl results,
where the first error comes from $|V_{ub}/V_{cb}|$ \cite{CKMF},
the second from the branching ratios (see Table \ref{tab:data}).
\begin{table}[tpb]
\renewcommand{\arraystretch}{1.1}
\begin{center}
\begin{tabular}{|c|c|c|c|}
\hline
$B^0\to\pi^+\pi^-$ & $B^+\to\pi^+\pi^0$ & $B^0\to\pi^0\pi^0$ & 
$A_{CP}(\pi^+\pi^0)$ \\
\hline
$4.6\pm 0.4$ & $5.3\pm 0.8$ & $1.90\pm 0.47$ &
$-0.07\pm 0.14$ \\
\hline\hline
$B^0\to K^+\pi^-$ & $B^+\to K^+\pi^0$ & $B^+\to K^0\pi^+$ & 
$B^0\to K^0\pi^0$\\
\hline
$18.1\pm 0.8$ & $12.8\pm 1.1$ & $19.6\pm 1.5$ &
$11.2\pm 1.4$\\
\hline\hline
$A_{CP}(K^+\pi^-)$ & $A_{CP}(K^+\pi^0)$ & $A_{CP}(K^0\pi^+)$ & 
$A_{CP}(K^0\pi^0)$\\
\hline
$-0.09\pm 0.03$ & $0.00\pm 0.07$ & $-0.01\pm 0.06$ &
$+0.03\pm 0.37$ \\ 
\hline\hline
$B^0\to K^+ K^-$ & $B^+\to K^+\bar K^0$ & $B^0\to K^0 \bar K^0$ &  \\
\hline
$ < 0.6$ & $<1.3$ & $<1.6$ & \\
\hline
\end{tabular}
\end{center}
\caption[]{\it Current world average values for $B\to \pi\pi$, $K\pi$
branching ratios (CP averaged, in units of $10^{-6}$) 
and direct CP asymmetries
$A_{CP}(f)$ $\equiv$ 
$(\Gamma(\bar B\to\bar f)-\Gamma(B\to f))/
(\Gamma(\bar B\to\bar f)+\Gamma(B\to f))$ \cite{LP03}.\label{tab:data}}
\end{table}
The parameter $\tilde r$ differs from $r$ in the numerator
through a different annihilation correction
($b_3$ instead of $b_3+2 b_4$ (\ref{rqcd})) and through
small $SU(3)$ breaking differences in the light-meson
distribution amplitudes ($\pi K$ instead of $\pi\pi$).
In the denominator $\tilde r$ has the pure tree amplitude $a_1+a_2$,
while $r$ has $a_1$ corrected by small penguin and annihilation terms.
Despite these differences in details, the structure of $r$ and $\tilde r$
are very similar. In fact, the theoretical value for $\tilde r$ from
QCD factorization 
\begin{equation}\label{rtilnum}
\tilde r=0.081\pm 0.016 \pm 0.016 =0.081 \pm 0.023
\end{equation}
is very close to the corresponding value for $r$,
and both are in agreement with the experimental number in (\ref{rtilde}).

A final comment concerns the branching ratio for $B\to\pi^0\pi^0$, which 
appears to be somewhat larger experimentally (Table \ref{tab:data})
than expected in recent theoretical calculations \cite{BN}, even though
the error bar is still large. It should be stressed that $B\to\pi^0\pi^0$,
being  colour-suppressed (amplitude involving $a_2$), 
is highly sensitive to the dynamics of
hard spectator interactions, which so far are only known to lowest order
in QCD and depend on poorly known input within the factorization framework.
These uncertainties strongly affect $B\to\pi^0\pi^0$, but
are considerably smaller in $B\to\pi^+\pi^0$ and $B\to\pi^+\pi^-$, as already
pointed out in \cite{BBNS1}. In \cite{BN} a scenario with an enhanced
$B\to\pi^0\pi^0$ rate, without the need for very unusual hadronic input,
was suggested. 
Such a scenario with large $a_2$ could be checked 
using the factorization test discussed in the preceding subsection.
We emphasize, however, that the uncertainties in $a_2$ specific to 
$B\to\pi^0\pi^0$
have essentially no impact on the penguin-to-tree ratio $r\,\exp(i\phi)$,
because the dominant hadronic physics is characteristically different.
Even a relatively large value of $B(B\to\pi^0\pi^0)$ does therefore not
invalidate the theoretical results for $r\,\exp(i\phi)$.

\subsection{Annihilation Decays}

Amplitudes from weak annihilation represent power
suppressed corrections, which are uncalculable in QCD
factorization and so far need to be estimated relying on
models \cite{BBNS3}. 
At present there are no indications that annihilation terms
would be anomalously large, but they do contribute to the theoretical 
uncertainty. Effectively, annihilation corrections may be considered
as part of the penguin amplitudes. To some extent, therefore, they
are tested with help of the quantity $\tilde r$ discussed in the
previous subsection. Nevertheless, in order to disentangle
their impact from other effects it is of great interest
to test annihilation separately. This can be done with decay
modes that proceed through annihilation or at least have a dominant
annihilation component.

An example is the pure annihilation channel $B_d\to D^-_s K^+$.
Even though this case is 
somewhat different from the reactions of primary interest here,
because of the charmed meson in the final state, it is still useful
to cross-check the typical size of annihilation expected in model
calculations. Treating the $D$ meson in the model estimate for annihilation
\cite{BBNS3} as suggested in \cite{BBNS2}, one finds a central
(CP-averaged) branching ratio of  $B(B_d\to D^-_s K^+)=1.2\times 10^{-5}$
and an upper limit of $5\times 10^{-5}$\cite{BCKMWS}. This is in agreement 
with the current experimental result $(3.8\pm 1.1)\times 10^{-5}$ 
(see refs. in \cite{BCKMWS}).

Additional tests should come from annihilation decays into
two light mesons, such as the $B\to KK$ modes in Table \ref{tab:data}.
These, however, are CKM suppressed and only upper limits are known
at present.
The $K^+\bar K^0$ and $K^0\bar K^0$ channels have both
annihilation and penguin contributions. On the other hand
$B\to K^+ K^-$ is a pure weak annihilation process and therefore
especially important. 
Further discussions can be found in \cite{BBNS3,BN}.


\section{Conclusions}\label{sec:concl}

In this paper we have proposed strategies to extract
information on weak phases from CP violation
observables in $B\to\pi^+\pi^-$ decays even in the presence
of hadronic contributions related to penguin amplitudes.
Our main results can be summarized as follows.

\begin{itemize}
\item
An efficient use of mixing-induced CP violation in $B\to\pi^+\pi^-$
decays, measured by $S$, can be made by combining it with the corresponding
observable from $B\to\psi K_S$, $\sin 2\beta$ or $\tau=\cot\beta$.
\begin{itemize}
\item
The unitarity triangle parameters $\bar\rho$ and $\bar\eta$ can then be 
obtained in closed form as functions of the observables
$\tau$, $S$ and the hadronic penguin parameters $r$, $\phi$
(see eqs. (\ref{rhotaueta}), (\ref{etataus})).
\item
The sensitivity on the hadronic quantities, which have typical
values $r\approx 0.1$, $\phi\approx 0.2$, is very weak.
In particular, there are no first-order corrections in $\phi$.
For moderate values of $\phi$ its effect is negligible.
\item
Neglecting $\phi$, the dependence of $\bar\eta$ on $r$ comes
merely through an overall factor $(1+r)$.  
The impact of the uncertainty in $r\approx 0.1$ becomes clearly visible
and is seen to be greatly reduced.
A simple determination of the unitarity triangle from $\tau$ and $S$
is thus possible (see eq. (\ref{etataus0})).
\end{itemize}
\item
The parameters $\bar\eta$, $1-\bar\rho$, $R_t$ and $\gamma$ are
bounded from below, depending only on $\tau$ and $S$ and essentially
without relying on hadronic input (see eq. (\ref{etabound})).
\item 
The parameter of direct CP violation $C$ depends much stronger on
hadronic input than $S$, but yields complementary information
and can constrain $r$ and $\phi$ within the standard model.
\item
An alternative analysis of $S$ and $C$ especially suitable for the presence 
of a new-physics phase in $B_d$--$\bar B_d$ mixing was discussed.
\end{itemize}

As an input to the phenomenological discussion we also studied
the calculation of the penguin parameters $r$ and $\phi$ in QCD:

\begin{itemize}
\item
We have analyzed $r$ and $\phi$ within QCD factorization
with a particular view on theoretical uncertainties.
\item
$B\to\pi^+\pi^-$ amplitudes can be expanded simultaneously 
in $1/m_b$ and $1/N$, which leads to an interesting pattern
of simplifications. All power corrections suffering from infrared
endpoint divergences in the factorization formalism are at least
of second order in this double expansion, while the most important
effects survive at linear oder in $1/m_b$ or $1/N$.
\item 
The different contributions to $B\to\pi^+\pi^-$
amplitudes, the tree-component, the penguin-to-tree ratio and
annihilation effects, appear in similar form in other $B$ decay
channels, such as $B^+\to\pi^+\pi^0$, $B^+\to\pi^+ K^0$ and $B\to K^+ K^-$. 
These can be used to validate theoretical predictions,
separately for the various components.

\end{itemize}

The results presented in this paper should be useful for interpreting the 
forthcoming experimental measurements of CP violation in $B\to\pi^+\pi^-$
decays in a transparent way and help to achieve a reliable control over
theoretical uncertainties.

\section*{Acknowledgements}

This work is supported in part by the Deutsche Forschungsgemeinschaft
(DFG) under contract BU 1391/1-2.

\vfill\eject

\end{document}